\documentclass[fleqn,usenatbib]{mnras}

\usepackage{newtxtext,newtxmath}
\usepackage[T1]{fontenc}
\DeclareRobustCommand{\VAN}[3]{#2}
\let\VANthebibliography\thebibliography
\def\thebibliography{\DeclareRobustCommand{\VAN}[3]{##3}\VANthebibliography}

\usepackage{graphicx}	
\usepackage{amsmath}

\title[A technique to precisely infer limb darkening]{A regularisation technique to precisely infer limb darkening using transit measurements: can we estimate stellar surface magnetic fields?}

\author[Verma et al.]{
Kuldeep Verma,$^{1}$\thanks{E-mail: kuldeep.phy@itbhu.ac.in (KV)}
Pierre F. L. Maxted,$^{2}$\thanks{E-mail: p.maxted@keele.ac.uk (PM)}
Anjali Singh,$^{1}$\thanks{E-mail: anjalisingh.rs.phy23@itbhu.ac.in (AS)}
H.-G. Ludwig,$^{3}$
Yashwardhan Sable $^{1}$
\\
$^{1}$Department of Physics, Indian Institute of Technology (BHU), Varanasi-221005, India\\
$^{2}$Astrophysics group, Keele University, Staffs, ST5 5BG, UK\\
$^{3}$Landessternwarte - Zentrum f\"{u}r Astronomie der Universit\"{a}t Heidelberg, K\"{o}nigstuhl 12, 69117 Heidelberg, Germany
}

\date{Accepted XXX. Received YYY; in original form ZZZ}
\pubyear{2024}

\begin{document}
\label{firstpage}
\pagerange{\pageref{firstpage}--\pageref{lastpage}}
\maketitle

\begin{abstract}
The high-precision measurements of exoplanet transit light curves that are now available contain information about the planet properties, their orbital parameters, and stellar limb darkening (LD). Recent 3D magneto-hydrodynamical (MHD) simulations of stellar atmospheres have shown that LD depends on the photospheric magnetic field, and hence its precise determination can be used to estimate the field strength. Among existing LD laws, the uses of the simplest ones may lead to biased inferences, whereas the uses of complex laws typically lead to a large degeneracy among the LD parameters. We have developed a novel approach in which we use a complex LD model but with second derivative regularisation during the fitting process. Regularisation controls the complexity of the model appropriately and reduces the degeneracy among LD parameters, thus resulting in precise inferences. The tests on simulated data suggest that our inferences are not only precise but also accurate. This technique is used to re-analyse 43 transit light curves measured by the NASA {\it Kepler} and TESS missions. Comparisons of our LD inferences with the corresponding literature values show good agreement, while the precisions of our measurements are better by up to a factor of 2. We find that 1D non-magnetic model atmospheres fail to reproduce the observations while 3D MHD simulations are qualitatively consistent. The LD measurements, together with MHD simulations, confirm that Kepler-17, WASP-18, and KELT-24 have relatively high magnetic fields ($>200$ G). This study paves the way for estimating the stellar surface magnetic field using the LD measurements.
\end{abstract}


\section{Introduction}
\label{sec:intro}
An accurate understanding of the stellar centre-to-limb variation (CLV) of specific intensity or the so-called limb darkening (LD) is necessary to interpret a variety of astronomical observations including measurements of the exoplanet transit light curves. This becomes particularly important in the contemporary era of space-based high-precision photometry, for example from the NASA {\it Kepler} \citep{boru10} and Transiting Exoplanet Survey Satellite \citep[TESS;][]{rick15}, and from the upcoming ESA PLAnetary Transits and Oscillations of stars telescope \citep[PLATO;][]{raue14,raue24}. The high-quality data delivered by the James Webb Space Telescope \citep[JWST;][]{gard06} and similar precision data expected from other upcoming facilities such as the Extremely Large Telescope \citep[ELT;][]{gilm07} and Atmospheric Remote-sensing Infrared Exoplanet Large-survey \citep[ARIEL;][]{tine18} demand better treatment of LD. Its inadequate treatment in the analysis of the transit light curves not only limits the precision with which LD itself can be characterised, but also hinders the precision of the inferred planet and orbital properties.  

Several limb darkening laws have been proposed and used in the literature. The popular ones are the linear law introduced by \citet{miln21}, the quadratic law \citep{kopa50}, the logarithmic law \citep{klin70}, the square-root law \citep{diaz92}, the power-2 law \citep{hest97}, the nonlinear or Claret 4-parameter law \citep{clar00} and the exponential law proposed by \citet{clar03}. These models of LD span a wide range in complexity, with the simplest linear law having only one free parameter and the most complex nonlinear law possessing four free parameters.

The impact of uncertainties in LD models on the determination of properties of the exoplanet system has been extensively studied during the last two decades \citep[see e.g.][just to name a few]{sing08,howa11,csiz13,mull13,espi16,more17,neil17,pate22}. In general, there is a consensus that the use of the linear law results in biased parameter values \citep[see e.g.][]{espi16}, suggesting that it is an overly simplistic model. Moreover, it is known that the quadratic law is a poor model for the LD profile of cool stars \citep{pate22} and results in biased parameters when fitting the transit light curves \citep{espi16,coul24}. This makes it difficult to interpret the LD coefficients obtained by fitting the transit light curves using a quadratic law \citep{howa11}. The power-2 law is a more realistic LD law \citep{more17} and was already used by \citet{maxt18} to characterise the accuracy of the LD profiles from the STAGGER-grid stellar model atmospheres. However, as discussed by \citet{maxt23}, it is not possible to investigate the constraints on the LD profile of real stars using a simple 1- or 2-parameter LD law by fitting the transit light curves because these simple laws impose strong assumptions on the shape of the LD profile. It is not possible to determine whether the shape of an assumed LD law is accurate unless we fit the light curve data using a more complicated LD law, such as the nonlinear law, which has the flexibility to deviate from the shapes assumed by the simpler LD laws. The nonlinear law provides unbiased parameter estimates \citep[see e.g.][hereafter M23]{maxt23}. However, as can be seen in Figure~6 of M23, there is a large degeneracy among the LD parameters. This could be due to an inappropriate functional form of the nonlinear law or because of the fact that this model is more complex than the underlying true CLV of stars or is a combined result of both. The large degeneracy not only hampers the precision of the inferred LD parameters, but also of the determined planetary and orbital properties, as they typically have finite correlations with the LD parameters.    

The CLV of the specific intensity depends on the physical condition near the stellar photosphere, and hence its precise measurement can provide us useful information. In fact, with the precision that M23 already determined the LD parameters, he observed an on average mild but persistent systematic offset between the observed LD profiles and those predicted by various models of the stellar atmosphere. He attributed this offset to the neglect of magnetic fields in the models. Recent studies with magneto-hydrodynamical (MHD) simulations of stellar atmospheres have shown that the CLV indeed depends on the mean magnetic field in the photosphere \citep{norr17,ludw23,kost24}. In particular, increasing the magnetic field leads to a relative brightening of the stellar limb.  

In this study, we have used the nonlinear LD model but with second derivative regularisation while fitting the model transit light curves to the observed data. This novel approach of fitting with regularisation allows us to control the complexity of the nonlinear model appropriately, thus providing inferences of the LD parameters, planet properties, and orbital parameters with unprecedented precision. We test our method on 900 simulated transit light curves and show that it provides accurate results. The technique is further used to re-analyse 43 high-quality exoplanet transit light curves measured by the {\it Kepler} and TESS missions and infer various properties, including the LD parameters, with unprecedented precision. Finally, we compare our measurements with the predictions of 1D non-magnetic model atmospheres as well as 3D MHD simulations. 

The paper is organised in the following order. We briefly describe our sample of observed stars in Section~\ref{sec:data}. The detailed method for fitting the exoplanet transit models to the observed light curve is developed in Section~\ref{sec:method}. In Section~\ref{sec:results}, we test our method and present the results. The conclusions of the paper are summarised in Section~\ref{sec:conc}.

\section{Transit data}
\label{sec:data}
We used the sample of 43 stars analysed in M23. For details on target selection and pre-processing of the observed transit light curves, we refer the reader to the above-mentioned paper. Briefly, the sample consists of 33 {\it Kepler} and 10 TESS targets. The {\it Kepler} targets were selected such that they had a signal-to-noise ratio greater than 500 for the transit signal, an orbital period less than 30 d, a transit impact parameter less than 0.8, and a host star effective temperature less than 7000 K. For the TESS targets, the selection was such that the host stars were brighter than the visual magnitude of 11.5 showing transits at least 0.5 per cent deep due to planets having an orbital period less than 10 d. We used the measured effective temperature $T_{\rm eff}$, surface gravity $\log g$, and metallicity $[{\rm Fe}/{\rm H}]$ for all the 43 stars from M23.

\section{Fitting method}
\label{sec:method}
We use the publicly available package \textsc{BATMAN}\footnote{https://lkreidberg.github.io/batman/docs/html/index.html} \citep[version 2.4.8,][]{krei15} to calculate the model transit light curves. This code has been extensively used in the literature and is well tested (see e.g. M23). \textsc{BATMAN} offers several options to users for the LD model. In this study, we used it with the nonlinear model \citep{clar00},
\begin{equation}
    I(\mu) = 1 - \sum_{i = 1}^4 a_i \left(1 - \mu^{i/2}\right),
    \label{eq:ld_profile}
\end{equation}
where the independent variable $\mu$ is the cosine of angle $\theta$ between the surface normal vector and the line of sight, i.e. $\mu = \cos(\theta)$, and the coefficients $a_1$, $a_2$, $a_3$ and $a_4$ are constants. $I(\mu)$ is the specific intensity normalised so that $I(1) = 1$.

We used a Bayesian framework based on the Markov Chain Monte Carlo (MCMC) methods to fit the model transit light curves to the observed ones. Since there are already several text books and reviews on Bayesian statistics \citep[see e.g.][]{jayn03,gelm13} and MCMC algorithms \citep[see e.g.][]{broo11,shar17,hogg18}, we describe them here only briefly and highlight the regularisation aspect.

In the context of parameter estimation, Bayes' theorem provides a way to update the model parameters in light of any newly acquired data. In other words, it enables the calculation of the posterior probability distribution (PPD) of the parameters given the new data,
\begin{equation}
    P({\bf \Theta}|{\bf D})=\frac{P({\bf \Theta}) P({\bf D}|{\bf \Theta})}{ P({\bf D})},
    \label{eq:bayes}
\end{equation}
where ${\bf D}$ represents the set of observations or the data, ${\bf \Theta}$ is the set of model parameters, $P({\bf D}|{\bf \Theta}) \coloneqq \mathscr{L}$ or the likelihood is the probability of observing the data given the model parameters, $P({\bf \Theta})$ or the prior is the probability distribution of the model parameters before seeing the new data, and $P({\bf D})$ or the evidence is the total probability of observing the data. The evidence is a normalisation constant and can be calculated by integrating the likelihood over all the model parameters. 

In this study, the observed flux as a function of time represents the data ${\bf D}$. Assuming the uncertainties on the flux values are independent and Gaussian distributed, we define the logarithm of the likelihood as,
\begin{eqnarray}
    \ln \mathscr{L} = &-& \frac{1}{2} \sum_{j = 1}^N \ln \left(2 \pi f^2 \sigma_j^2 \right) - \frac{1}{2} \sum_{j = 1}^N \left(\frac{F_{{\rm obs},j} - F_{{\rm mod},j}}{f \sigma_j} \right)^2\nonumber\\
    &-& \frac{1}{2} \lambda^2 \sum_{j = 1}^{100} \left[\frac{d^2 I(\mu)}{d\mu^2}\right]_{\mu = \mu_j}^2.
    \label{eq:likelihood}
\end{eqnarray}
The first two terms involve summation over the number of data points $N$ in the flux time-series, and they together represent the standard logarithm of the likelihood. The quantities $F_{{\rm obs},j}$ and $\sigma_j$ are the measured value of the flux and the corresponding observational uncertainty, respectively, whereas the quantity $F_{{\rm mod},j}$ is the model flux value. We assume that the uncertainties $\sigma_j$ are well estimated, however they can be wrong by a constant factor $f$. This factor $f$ is treated as a free parameter in the fitting process. The last term represents a second derivative regularisation and is proportional to the sum of squares of the second derivative of $I(\mu)$, 
\begin{equation}
    \frac{d^2 I(\mu)}{d\mu^2} = \frac{1}{4} \sum_{i = 1}^4 i (i - 2) a_i \mu^{(i - 4)/2},
    \label{eq:ld_profile_2}
\end{equation} 
calculated at 100 uniformly spaced $\mu$ values between 0.01 and 1. The proportionality constant $\lambda$ is called the regularisation parameter. It should be noted that $\lambda$ is dimensionless because $I(\mu)$ is normalised and has no dimension. The regularisation term penalises large curvatures in the LD profile to an extent that is determined by the value of $\lambda$ (see Section~\ref{subsec:lambda}). We tried using the first derivative regularisation as well, but the preliminary results indicated substantially worse performance compared to the second derivative regularisation.    

\begin{figure}
	\includegraphics[width=\columnwidth]{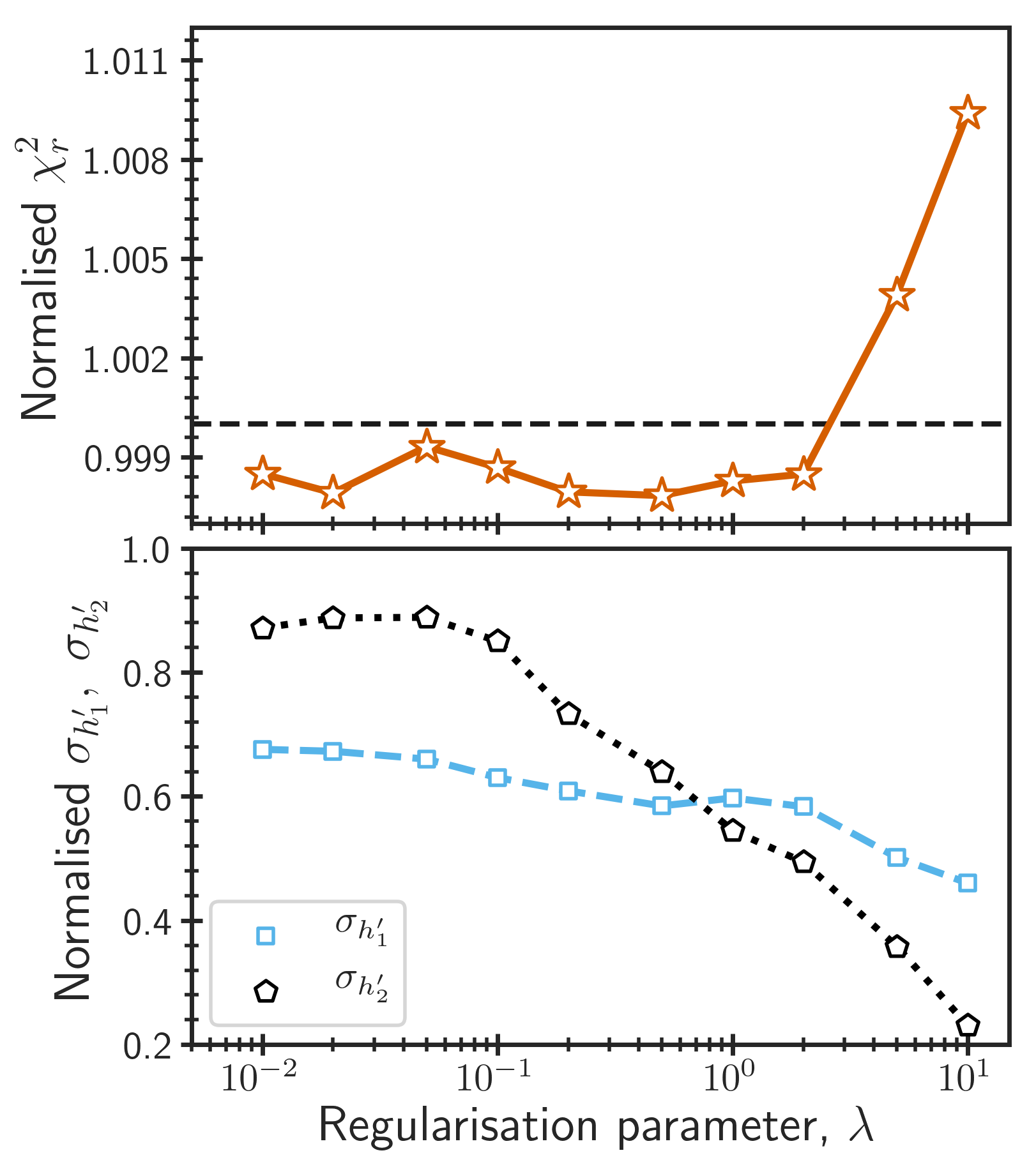}
    \caption{Reduced chi-square (top panel) and uncertainties on $h_1'$ and $h_2'$ (bottom panel) obtained by fitting the transit light curve of Kepler-5 as a function of the regularisation parameter. The $\chi_r^2$ and uncertainties are normalised such that they are 1 for the fit with $\lambda = 0$. The dashed horizontal line in the top panel marks the normalised $\chi_r^2$ value of 1. As also shown in the legend, the square and pentagon symbols in the bottom panel represent the normalised uncertainties on $h_1'$ and $h_2'$, respectively. The points are connected with lines to guide the eye.}
    \label{fig:lamda}
\end{figure}

Our transit model has a set of 10 free parameters; these are the orbital period $P$, the mid-transit time $T_0$, the planet-to-star radius ratio $R_p/R_*$, the ratio of the host star radius and the orbital semi-major axis $R_*/a$, the transit impact parameter $b$, the LD parameters $\{a_1, a_2, a_3, a_4\}$, and the logarithm of the error scale factor $f$. For all of them, we used uniform priors with $P \in [0, 500]$ d, $T_0 \in [0, 2460300]$ d (the upper Barycentric Julian Date corresponds to a calendar date of December 21, 2023, and hence the prior covers the full time span of all the observed transit light curves analysed in this study), $R_p/R_*$ and $R_*/a$ $\in [0, 0.5]$, $b \in [-1, 1]$, all the LD parameters $a_i$ $\in [-50, 50]$ and $\log f \in [-0.5, 0.5]$. Note that, in principle, $T_0$ can have multiple values separated by $P$; in practice, however, the sampler finds a value closest to its initial guess (and the PPD remains uni-modal). 

Finally, we sampled PPD as defined in equation~(\ref{eq:bayes}) using the publicly available affine invariant MCMC ensemble sampler, \textsc{emcee}: the MCMC hammer\footnote{https://emcee.readthedocs.io/en/stable/} \citep{good10,fore13}. We used 100 walkers, a total of 5000 steps, and 4000 burn-in steps. The convergence of chains was confirmed by visually inspecting the trends in the parameter values and variances as a function of step number after the burn-in phase.

\begin{figure}
	\includegraphics[width=\columnwidth]{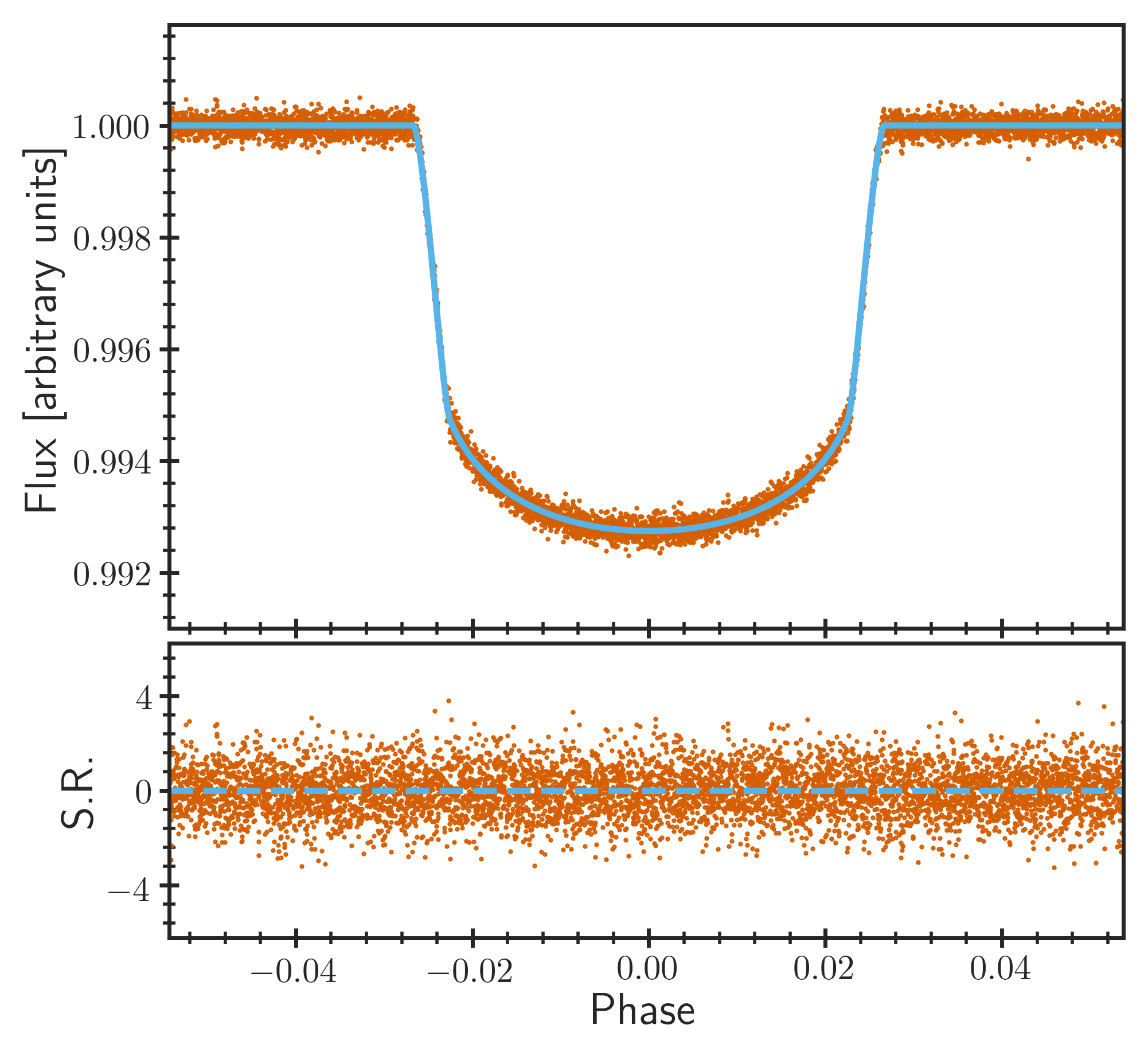}
    \caption{Phase folded light curve for Kepler-5 (top panel) and standardised residual (bottom panel). In the top panel, the dots show the observed data while the continuous curve represents the best-fitting model obtained with $\lambda = 0.2$. In the bottom panel, the standardised residual is the residual divided by the corresponding observational uncertainty. The dashed horizontal line marks the zero level.}
    \label{fig:phase_diagram}
\end{figure}

\begin{figure*}
	\includegraphics[width=\textwidth]{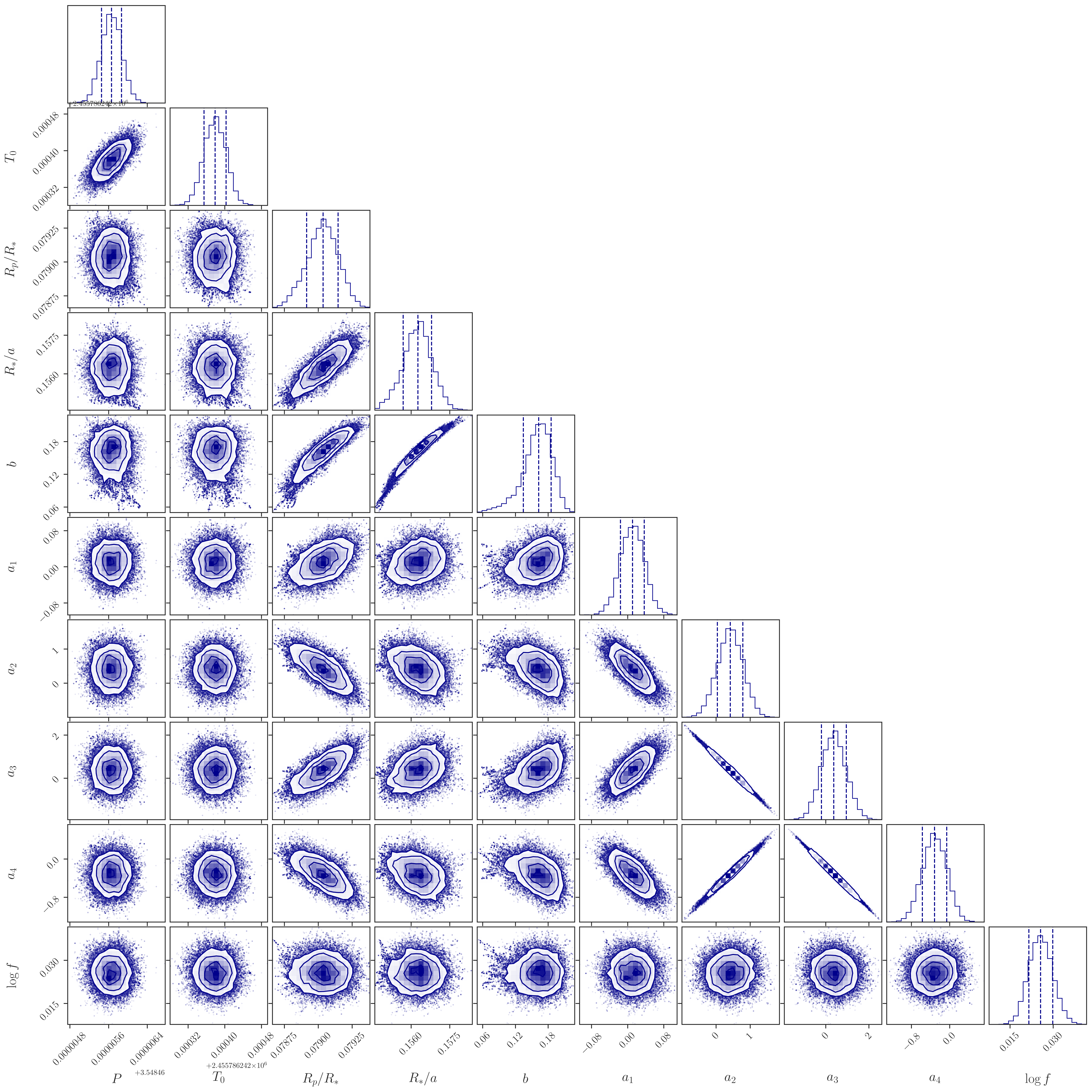}
    \caption{Corner diagram showing the posterior probability distribution resulting from the fit with $\lambda = 0.2$ to the transit light curve of Kepler-5.}
    \label{fig:corner}
\end{figure*}

\subsection{Value of regularisation parameter}
\label{subsec:lambda}
Before we can use the above method, we must determine the penalty for the large $I(\mu)$ curvature by selecting an appropriate value for the regularisation parameter. From the definition of the likelihood, a small or close to zero value of $\lambda$ would clearly mean that all LD coefficients are free to vary during the fitting process. In such a case, the four free parameters in the nonlinear law provide a large flexibility in the model, and hence it reproduces the observed high-precision photometric data quite well. However, if it turns out that the true LD profiles of stars are simpler than those predicted by the nonlinear law, then its use would lead to overfitting of the data. The large degeneracy among the LD parameters seen in M23 could potentially be a result of this. On the other hand, it may be noted from equation~\ref{eq:likelihood} that a choice of an extremely large value of $\lambda$ during the fitting process would lead to small values of the second derivative of $I(\mu)$ for all values of $\mu$, i.e., 
\begin{equation}
    - a_1 \mu^{-3/2} + 3 a_3 \mu^{-1/2} + 8 a_4 \approx 0.
    \label{eq:constraint}
\end{equation}
Such a choice of $\lambda$ would lead to a simpler LD law with only effectively three free parameters (since $a_1$, $a_3$ and $a_4$ satisfy the above constraint) and could lead to underfitting of data. We follow the procedure below to choose $\lambda$ in a way that avoids both overfitting and underfitting the data.

\begin{figure}
	\includegraphics[width=\columnwidth]{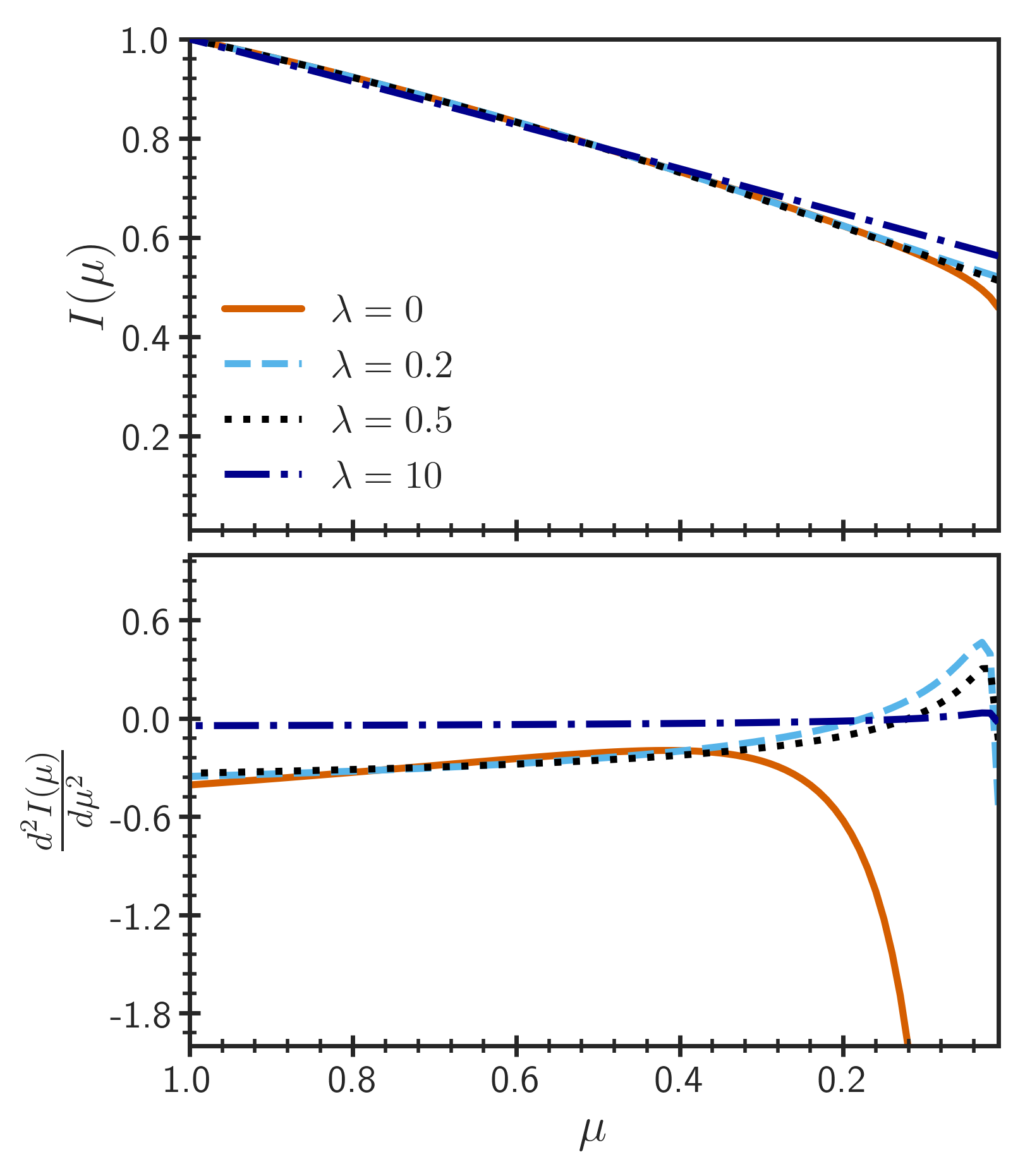}
    \caption{Limb darkening profiles obtained by fitting the observed transit light curve of Kepler-5 with different values of the regularisation parameter (top panel) and their second derivative (bottom panel). As indicated in the legend, the solid, dashed, dotted, and dash-dotted curves represent profiles obtained with $\lambda = 0$, $0.2$, $0.5$, and $10$, respectively. In the bottom panel, the $y$-axis is restricted in the range from -2 to 1 for clarity.}
    \label{fig:ld_profile}
\end{figure}

To determine the optimal value of $\lambda$, we performed 11 different fits to the observed transit light curve of Kepler-5 with $\lambda = \{0, 0.01, 0.02, 0.05, 0.1, 0.2, 0.5, 1, 2, 5, 10\}$, and plotted the goodness of fit as defined by the reduced chi-square, 
\begin{equation}
    \chi_r^2 = \frac{1}{N - 10} \sum_{j = 1}^N \left(\frac{F_{{\rm obs},j} - F_{{\rm mod},j}}{\sigma_j} \right)^2,
    \label{eq:chi2r}
\end{equation}
as a function of $\lambda$ in Figure~\ref{fig:lamda}. Note that in the above fits, except for $\lambda$, everything else, including the priors on the model parameters and the initial guesses for the different walkers, remains unchanged. The values of $\chi_r^2$ are normalised such that it is 1 for the fit with $\lambda = 0$. At low values of $\lambda$, the normalised $\chi_r^2$ shows small fluctuations. This is likely a result of small differences in the convergence of MCMC chains for fits with different values of $\lambda$. However, the normalised $\chi_r^2$ starts to increase monotonically for $\lambda > 0.5$, indicating that the LD model is becoming increasingly simpler and inadequate to model the data (i.e. underfitting). We wish to emphasise that the increasing profile at large $\lambda$ values is robust. The above suggests that, as far as the goodness of fit is concerned, any value of $\lambda \le 0.5$ is a reasonable choice.

In addition to the goodness of fit, we also noted the precision with which we inferred the LD profile as a function of $\lambda$. Since the LD parameters are strongly correlated, it is useful to introduce new sets of variables that have relatively low correlations \citep[see e.g.][]{maxt18,maxt23}. Following M23, we use the parameters,
\begin{equation}
    h_1' = I(\mu = 2/3)
    \label{eq:h1p}
\end{equation}
and
\begin{equation}
    h_2' = h_1' - I(\mu = 1/3),
    \label{eq:h2p}
\end{equation}
to characterise the LD profile. Given the LD coefficients $a_i$, both $h_1'$ and $h_2'$ can easily be calculated using equations~\ref{eq:ld_profile}, \ref{eq:h1p} and \ref{eq:h2p}. In the bottom panel of Figure~\ref{fig:lamda}, we show the uncertainties in $h_1'$ and $h_2'$ obtained from PPDs as a function of $\lambda$. The uncertainties are normalised such that they are 1 for the fit with $\lambda = 0$. The uncertainties in both $h_1'$ and $h_2'$ decrease with an increasing value of $\lambda$. This is expected as a larger value of $\lambda$ leads to tighter constraints on the LD profile. 

Based on the above results, it is tempting to choose $\lambda = 0.5$ as the optimal value because it keeps $\chi_r^2$ at a low level and maximises the precision of inferences of $h_1'$ and $h_2'$ for $\lambda \le 0.5$. However, we wish to point out that there is a small chance that $\lambda = 0.5$ is already oversimplifying the LD model. To avoid potential biases in our inferences due to the oversimplification of the LD model, we take a conservative approach and give up the precision of the LD measurements slightly and choose $\lambda = 0.2$ as the optimal or reference value and use it from now on (unless mentioned otherwise). This choice provides a good fit to the data, as seen in Figure~\ref{fig:phase_diagram}, and results in reasonably precise LD parameters. As we can see in the corner plot \citep{fore16} in Figure~\ref{fig:corner}, the distributions are uni-modal, and the model parameters are well determined. Furthermore, the degeneracy among the LD parameters is smaller compared to that found in M23 (see their Figure~6; particularly, compare the panels corresponding to $a_1$ and $a_2$, $a_1$ and $a_3$, and $a_1$ and $a_4$). Although there is still substantial degeneracy among these parameters (see, particularly, the panels corresponding to $a_2$ and $a_3$, $a_2$ and $a_4$, and $a_3$ and $a_4$), the brightness distribution of the projected stellar disk is well constrained, resulting in precise determinations of the LD parameters $h_1'$ and $h_2'$.

The top panel of Figure~\ref{fig:ld_profile} presents Kepler-5 LD profiles obtained by fitting its transit light curve with four different values of $\lambda$. Clearly, all the profiles look similar for $\mu \gtrapprox 0.2$, indicating that this part of the profile is well constrained by the observational data. On the other hand, regularisation has a significant impact on the part of the profile close to the limb ($\mu \lessapprox 0.2$), where it systematically pushes the profile upward. The bottom panel shows the curvature profile. As expected, the LD profile obtained with $\lambda = 0$ has large curvatures (magnitude is larger than 2 near the limb), while the one found with $\lambda = 10$ has curvatures close to zero. The implications of choosing a value of $\lambda$ in the vicinity of 0.2 are discussed in Section~\ref{app:lambda}.

\begin{table*}
	\centering
	\caption{Inferred parameters for Kepler-5 from M23 and this work (V24 and V24test, see the related texts for details).}
	\label{tab:kepler-5}
	\begin{tabular}{lcccccccr}
		\hline
		Work & $P$ [d] & $h_1'$ & $h_2'$ & $C(h_1', h_2')$ & $R_p/R_*$ & $R_*/a$ & $b$ & $f$\\
		\hline
		  M23     & 3.55 & $0.863(1)$ & $0.168(5)$ & $-0.15$ & $0.0789(2)$ & $0.1563(6)$ & $0.15(3)$ & $1.06$ \\
        V24     & 3.55 & $0.864(2)$ & $0.169(5)$ & $-0.42$ & $0.0790(4)$ & $0.1564(7)$ & $0.16(4)$ & $1.06$ \\
        V24test & 3.55 & $0.863(1)$ & $0.168(4)$ & $-0.19$ & $0.0789(2)$ & $0.1564(6)$ & $0.16(3)$ & $1.06$ \\
        \hline
	\end{tabular}
\end{table*}

\begin{figure}
	\includegraphics[width=\columnwidth]{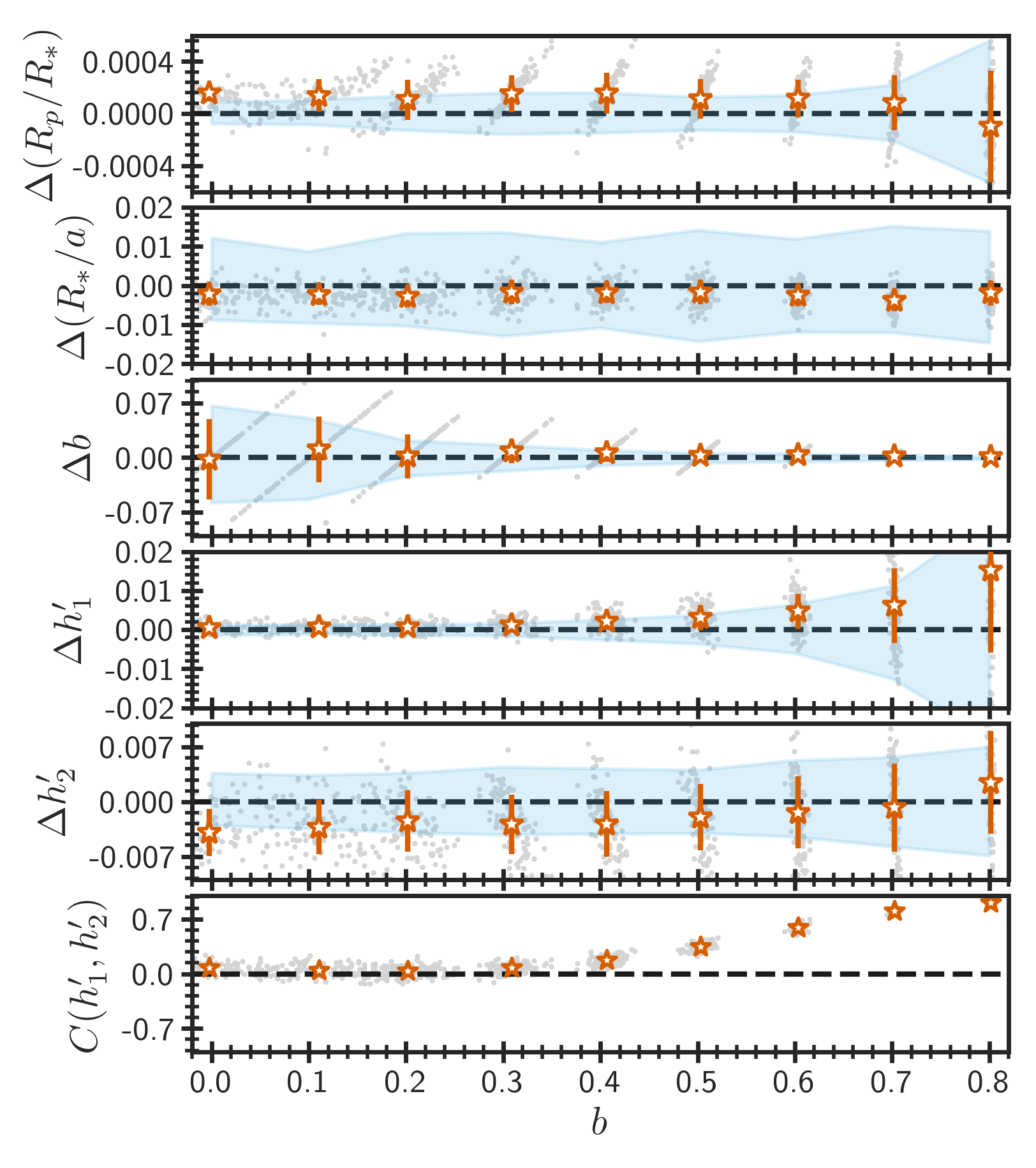}
    \caption{Orbital and limb darkening parameters recovered from our fits performed with $\lambda = 0.2$ to 900 simulated light curves of a transiting hot Jupiter. The grey dots show the differences between the fitted parameter values and the corresponding true values. For clarity, we have not included errorbars on the dots. However, the shaded regions show the uncertainties calculated from the PPDs of nine simulations with different $b$ values and can be considered representative of the errorbars on the individual dots. The points with errorbar show differences between the mean parameter values obtained from the results of 100 simulations and the corresponding true values as a function of mean $b$ values calculated using 100 simulations. The errorbars are the standard deviations computed from the results of 100 simulations. In the bottom panel, the points show correlations between $h_1'$ and $h_2'$ computed from the PPDs. The horizontal dashed lines correspond to perfect recovery in the top five panels, while in the bottom panel, it marks zero correlation.}
    \label{fig:simulation}
\end{figure}

\section{Results}
\label{sec:results}
To ensure clarity, we begin this section by briefly describing its content. Before applying our fitting technique to the sample of stars discussed in Section~\ref{sec:data}, we test our method against both the observed and simulated data in Section~\ref{subsec:simulation}. Subsequently, all the stars in our sample are analysed, and the results are compared with that of M23 in Section~\ref{subsec:m23}. Finally, we compare our measurements of limb darkening with the predictions of 1D stellar atmospheric models as well as 3D MHD simulations in Section~\ref{subsec:models}.

\subsection{Tests on the observed and simulated data}
\label{subsec:simulation}
Our fitting method, when used with $\lambda = 0$, should provide results very similar (if not identical) to those obtained by M23. To test whether this is indeed the case, we have listed the parameters for Kepler-5 found with $\lambda = 0$ in Table~\ref{tab:kepler-5} (Work V24). The corresponding results from Table~2 of M23 are also provided in the table for the reader's convenience. At a glance, the agreement may appear reasonable, however a few discrepancies can be noticed with more careful inspection. For instance, our uncertainties on $h_1'$ and $R_p/R_*$ are about twice as large as those found in M23. Furthermore, the magnitude of our correlation between $h_1'$ and $h_2'$ is significantly larger than that of M23. We investigated these discrepancies in detail and found that they were the result of additional constraints used by M23. For every set of predicted LD coefficients \{$a_1$, $a_2$, $a_3$, $a_4$\} during the fitting process, he calculated the LD profile at 100 uniformly spaced $\mu$ values between 0.01 and 1 and rejected the solution if the coefficients did not correspond to a profile with $0 < I(\mu) < 1$ and $\frac{d I(\mu)}{d\mu} > 0$ for all values of $\mu$. We performed a test with these additional constraints; the resulting parameters are also listed in the table (Work V24test). Clearly, including these constraints eliminates the discrepancies almost completely. However, since recent MHD simulations with high surface magnetic fields predict LD profiles that are neither monotonic nor limited to $0 < I(\mu) < 1$ \citep{ludw23}, we did not consider these constraints in the subsequent analysis.

To thoroughly evaluate the accuracy of our inferences, we performed a test on simulated data. We generated a set of 900 simulated transit light curves for 9 uniformly spaced $b$ values between 0 and 0.8 using \textsc{BATMAN}. In simulations, we used $P = 3.5$ d, $R_p/R_* = 0.08$, $R_*/a = 0.15$ and the solar LD profile computed by \citet{kost22}. The light curves were sampled at 2000 points uniformly distributed throughout the duration of the transit. For each value of $b$, we generated 100 light curves assuming Gaussian random noise in the flux with a standard deviation of 100 ppm. This is similar to the signal-to-noise in the {\it Kepler} light curves of moderately bright stars like Kepler-5 with transiting hot Jupiters.

We fitted all the 900 simulated light curves with the reference value of $\lambda = 0.2$. The fitted parameter values are compared with the corresponding true values in Figure~\ref{fig:simulation}. To clearly observe biases and trends, we use the results of 100 simulations for each $b$ value to compute the mean and standard deviation of the parameters and compare them with the corresponding true values. As we can see in the figure, all parameters are recovered reasonably well with systematic offsets generally within the statistical uncertainties. This demonstrates that our inferences are not only precise, but also accurate. Note that, except for $R_*/a$, the uncertainties obtained from the PPDs are consistent with the Monte Carlo (MC) errorbars. The large uncertainties on $R_*/a$ from PPDs compared to MC estimates are due to its strong anti-correlation with $P$. Note that the simulated light curves contain only one transit and, hence, $P$ is not well constrained. The light curves like the one of Kepler-5 having multiple transits constrain $P$ well, and hence, the resulting uncertainty on $R_*/a$ from PPD is substantially smaller (see Figure~\ref{fig:corner}). As can be seen in the figure, the correlation between $h_1'$ and $h_2'$ remains small for $b < 0.5$ and gradually increases beyond 0.5. 

\begin{table*}
	\centering
	\caption{Inferred parameters for all the 43 targets in our sample.}
	\label{tab:fit_params}
	\begin{tabular}{lrrrcrrrrr}
		\hline
		Star & $P$ [d] & $h_1'$ & $h_2'$ & $C(h_1', h_2')$ & $R_p/R_*$ & $R_*/a$ & $b$ & $\log f$ & $\chi^2_r$\\
		\hline
		  HAT-P-7 & 2.204735427(21) & $0.85862(97)$ & $0.17616(85)$ & $-0.08$ & $0.077472(29)$ & $0.24057(19)$ & $0.4924(16)$ & $0.1148(33)$ & $1.70$  \\
        Kepler-4 & 3.21367067(83) & $0.8494(96)$ & 
          $0.223(10)$ & $\phantom{-}0.51$ & $0.02468(25)$ & $0.1667(78)$ & $0.34(14)$ & $0.0383(48)$ & $1.20$  \\
        Kepler-5 & 3.54846566(21) & $0.8637(11)$ & 
          $0.1681(36)$ & $\phantom{-}0.02$ & $0.07904(12)$ & $0.15626(57)$ & $0.163(28)$ & $0.0257(42)$ & $1.13$   \\
        Kepler-6 & 3.23469935(13) & $0.8229(10)$ & 
          $0.2092(41)$ & $-0.20$ &  $0.09174(17)$ & $0.13391(46)$ & $0.167(27)$ & $0.0181(55)$ & $1.09$   \\
        Kepler-7 & 4.88548819(76) & $0.8548(46)$ & 
          $0.1894(38)$ & $\phantom{-}0.43$ & $0.08236(12)$ & $0.15051(48)$ & $0.5587(50)$ & $0.0333(51)$ & $1.17$   \\
        Kepler-8 & 3.52249824(22) & $0.864(15)$ & 
          $0.1796(69)$ & $\phantom{-}0.85$ & $0.09458(28)$ & $0.14658(39)$ & $0.7217(25)$ & $0.0221(46)$ & $1.11$   \\
        Kepler-12 & 4.43796264(22) & $0.84450(83)$ & 
          $0.1864(42)$ & $-0.16$ & $0.11779(16)$ & $0.12493(24)$ & $0.177(14)$ & $0.0612(43)$ & $1.33$   \\
        Kepler-14 & 6.79012279(95) & $0.8534(71)$ & 
          $0.1707(38)$ & $\phantom{-}0.70$ & $0.045525(90)$ & $0.13552(93)$ & $0.5974(82)$ & $0.0447(35)$ & $1.23$   \\ 
        Kepler-15 & 4.9427832(11) & $0.831(13)$ & 
          $0.2148(98)$ & $\phantom{-}0.83$ & $0.10273(38)$ & $0.10156(39)$ & $0.6826(44)$ & $0.0431(70)$ & $1.22$   \\ 
        Kepler-17 & 1.485710958(35) & $0.8406(12)$ & 
          $0.1759(49)$ & $-0.06$ & $0.13268(21)$ & $0.17651(38)$ & $0.169(17)$ & $0.1357(59)$ & $1.87$   \\
        Kepler-41 & 1.85555826(52) & $0.834(15)$ & 
          $0.208(11)$ & $\phantom{-}0.83$ & $0.10056(42)$ & $0.19611(92)$ & $0.6828(52)$ & $0.0200(91)$ & $1.10$   \\ 
        Kepler-43 & 3.02409225(15) & $0.8588(94)$ & 
          $0.2026(58)$ & $\phantom{-}0.70$ & $0.08557(19)$ & $0.14448(52)$ & $0.6584(42)$ & $0.0186(51)$ & $1.09$   \\ 
        Kepler-44 & 3.2467279(30) & $0.825(22)$ & 
          $0.201(19)$ & $\phantom{-}0.77$ & $0.08073(70)$ & $0.1436(24)$ & $0.642(19)$ & $0.0368(84)$ & $1.19$  \\
        Kepler-45 & 2.45524074(12) & $0.8169(83)$ & 
          $0.231(17)$ & $\phantom{-}0.70$ & $0.18176(82)$ & $0.09307(27)$ & $0.5607(62)$ & $0.0354(63)$ & $1.18$  \\
        Kepler-74 & 7.3407074(59) & $0.829(29)$ & 
          $0.179(19)$ & $\phantom{-}0.81$ & $0.09120(75)$ & $0.06555(75)$ & $0.701(11)$ & $0.0999(82)$ & $1.59$ \\
        Kepler-77 & 3.57878140(76) & $0.8200(30)$ & 
          $0.2107(69)$ & $\phantom{-}0.22$ & $0.09811(28)$ & $0.10348(59)$ & $0.366(17)$ & $0.0377(74)$ & $1.19$  \\ 
        Kepler-412 & 1.72086143(51) & $0.815(36)$ & 
          $0.197(12)$ & $\phantom{-}0.87$ & $0.10372(90)$ & $0.2052(10)$ & $0.7952(28)$ & $0.018(10)$ & $1.09$  \\
        Kepler-422 & 7.89144780(56) & $0.8361(34)$ & 
          $0.1969(46)$ & $\phantom{-}0.28$ & $0.09584(17)$ & $0.07359(23)$ & $0.4924(62)$ & $0.0462(38)$ & $1.24$  \\
        Kepler-423 & 2.68432839(13) & $0.8301(18)$ & 
          $0.2063(60)$ & $-0.01$ & $0.12395(26)$ & $0.12310(35)$ & $0.332(10)$ & $0.0336(73)$ & $1.17$  \\
        Kepler-425 & 3.7970169(13) & $0.802(11)$ & 
          $0.224(16)$ & $\phantom{-}0.77$ & $0.11440(70)$ & $0.08612(61)$ & $0.600(12)$ & $0.0252(94)$ & $1.13$  \\
        Kepler-426 & 3.21751888(38) & $0.856(28)$ & 
          $0.216(18)$ & $\phantom{-}0.83$ & $0.11827(77)$ & $0.10443(59)$ & $0.7254(53)$ & $0.0389(90)$ & $1.20$  \\
        Kepler-427 & 10.2910122(83) & $0.8272(45)$ & 
          $0.189(13)$ & $\phantom{-}0.38$ & $0.08989(64)$ & $0.0506(10)$ & $0.16(20)$ & $0.1721(67)$ & $2.21$  \\
        Kepler-433 & 5.3340855(79) & $0.8602(61)$ & 
          $0.172(12)$ & $\phantom{-}0.39$ & $0.06332(21)$ & $0.1433(20)$ & $0.05(17)$ & $0.0592(57)$ & $1.34$  \\
        Kepler-435 & 8.6001613(48) & $0.8628(66)$ & 
          $0.1881(93)$ & $\phantom{-}0.53$ & $0.06277(24)$ & $0.1400(19)$ & $0.422(31)$ & $0.0589(54)$ & $1.31$  \\
        Kepler-470 & 24.669387(67) & $0.8556(91)$ & 
          $0.163(13)$ & $\phantom{-}0.62$ & $0.08042(36)$ & $0.03733(51)$ & $0.429(29)$ & $0.2200(75)$ & $2.76$  \\
        Kepler-471 & 5.01423612(89) & $0.8701(75)$ & 
          $0.166(12)$ & $\phantom{-}0.59$ & $0.07658(31)$ & $0.1215(16)$ & $0.415(30)$ & $0.0819(75)$ & $1.46$  \\
        Kepler-485 & 3.24325949(32) & $0.8318(27)$ & 
          $0.1890(94)$ & $\phantom{-}0.14$ & $0.11795(42)$ & $0.11127(77)$ & $0.196(46)$ & $0.0303(71)$ & $1.15$  \\
        Kepler-489 & 17.276296(19) & $0.7873(59)$ & 
          $0.197(15)$ & $\phantom{-}0.49$ & $0.09207(73)$ & $0.02797(47)$ & $0.250(78)$ & $0.1098(83)$ & $1.66$  \\
        Kepler-490 & 3.2686961(15) & $0.8483(38)$ & 
          $0.196(10)$ & $\phantom{-}0.26$ & $0.09265(37)$ & $0.1305(13)$ & $0.265(45)$ & $0.0202(78)$ & $1.10$  \\
        Kepler-491 & 4.2253822(18) & $0.8180(57)$ & 
          $0.229(10)$ & $\phantom{-}0.45$ & $0.08045(37)$ & $0.0903(10)$ & $0.449(23)$ & $0.0330(84)$ & $1.17$  \\
        Kepler-492 & 11.720060(12) & $0.818(32)$ & 
          $0.193(19)$ & $\phantom{-}0.79$ & $0.09713(82)$ & $0.04003(47)$ & $0.699(10)$ & $0.2170(81)$ & $2.72$  \\
        Kepler-670 & 2.81650436(40) & $0.8205(43)$ & 
          $0.208(12)$ & $\phantom{-}0.40$ & $0.11981(51)$ & $0.11337(74)$ & $0.413(18)$ & $0.0154(72)$ & $1.08$  \\
        TrES-2 & 2.470613351(19) & $0.921(34)$ & 
          $0.2079(60)$ & $\phantom{-}0.85$ & $0.12379(75)$ & $0.12767(47)$ & $0.8469(11)$ & $0.0711(57)$ & $1.60$  \\
        HD 271181 & 4.2311147(14) & $0.9031(89)$ & 
          $0.162(16)$ & $\phantom{-}0.44$ & $0.08090(39)$ & $0.1305(25)$ & $0.327(65)$ & $0.0059(20)$ & $1.03$  \\
        KELT-23 & 2.25528764(11) & $0.8658(63)$ & 
          $0.1624(90)$ & $\phantom{-}0.68$ & $0.13299(30)$ & $0.13169(46)$ & $0.5283(63)$ & $0.0004(27)$ & $1.00$  \\
        KELT-24 & 5.55149335(53) & $0.8937(22)$ & 
          $0.1434(60)$ & $\phantom{-}0.31$ & $0.08706(13)$ & $0.09348(49)$ & $0.00(11)$ & $0.0126(32)$ & $1.07$  \\
        TOI-1181 & 2.10319351(46) & $0.8784(54)$ & 
          $0.160(11)$ & $\phantom{-}0.46$ & $0.07679(30)$ & $0.2465(31)$ & $0.321(44)$ & $0.0061(18)$ & $1.03$  \\
        TOI-1268 & 8.1577289(26) & $0.8549(84)$ & 
          $0.209(17)$ & $\phantom{-}0.29$ & $0.08958(65)$ & $0.0580(13)$ & $0.05(23)$ & $0.0145(50)$ & $1.09$  \\
        TOI-1296 & 3.9443737(18) & $0.8468(92)$ & 
          $0.173(19)$ & $\phantom{-}0.48$ & $0.07626(63)$ & $0.1545(41)$ & $0.28(12)$ & $0.0090(25)$ & $1.04$  \\
        WASP-18 & 0.941452432(59) & $0.8781(36)$ & 
          $0.1469(69)$ & $\phantom{-}0.48$ & $0.09738(19)$ & $0.2885(15)$ & $0.383(15)$ & $0.0006(26)$ & $1.00$  \\
        WASP-62 & 4.41193825(28) & $0.8881(19)$ & 
          $0.1433(61)$ & $\phantom{-}0.17$ & $0.11085(19)$ & $0.10312(43)$ & $0.243(20)$ & $0.0032(18)$ & $1.02$  \\
        WASP-100 & 2.84938185(34) & $0.892(11)$ & 
          $0.1347(92)$ & $\phantom{-}0.76$ & $0.08269(23)$ & $0.1871(14)$ & $0.571(10)$ & $0.0053(14)$ & $1.02$  \\
        WASP-126 & 3.28878666(48) & $0.8631(44)$ & 
          $0.197(11)$ & $\phantom{-}0.26$ & $0.07718(27)$ & $0.1271(14)$ & $0.05(16)$ & $0.0051(15)$ & $1.02$  \\
	\hline
	  \end{tabular}
\end{table*}

\begin{figure*}
	\includegraphics[width=\textwidth]{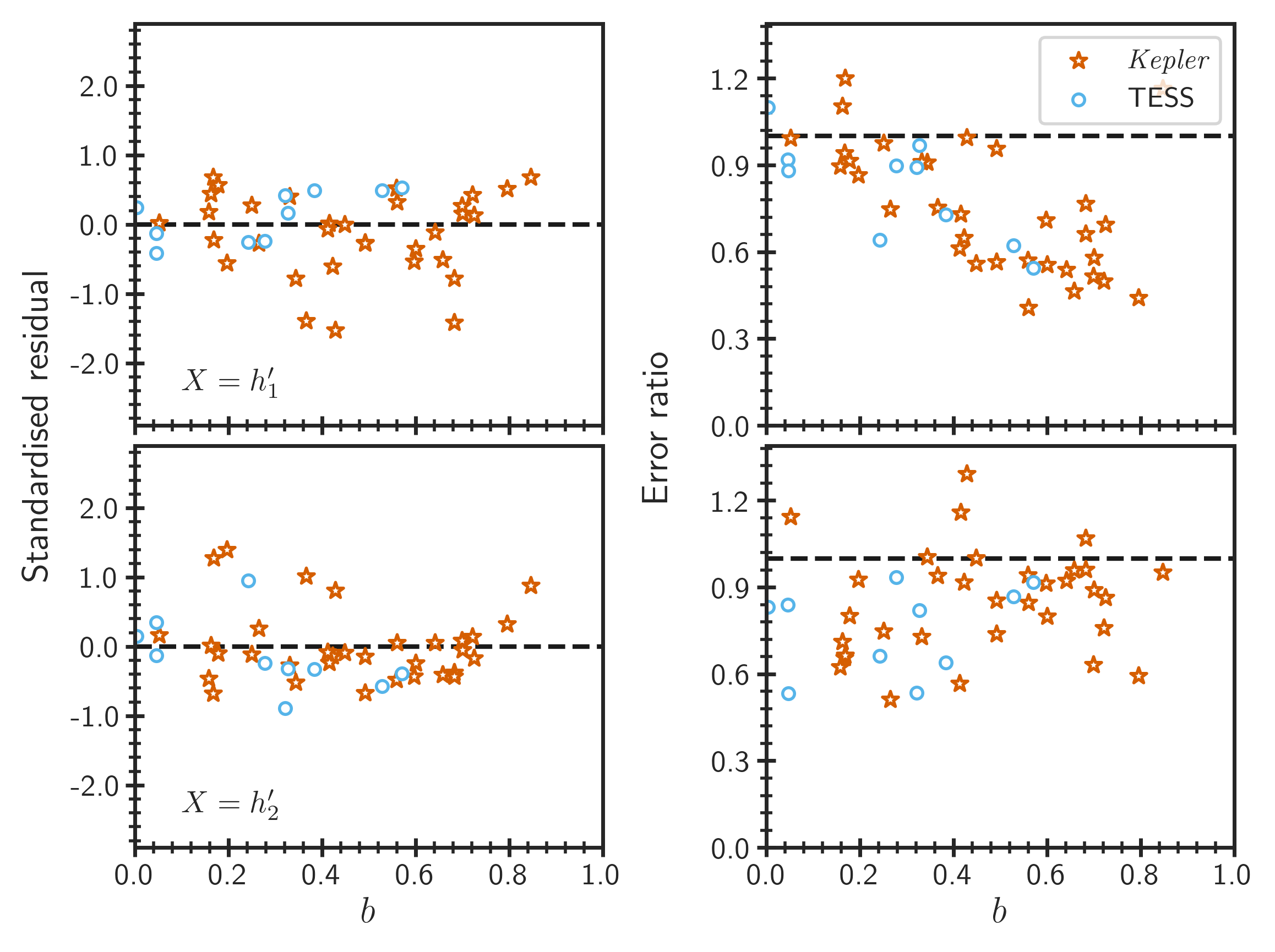}
    \caption{Comparison of our measurements of $h_1'$ and $h_2'$ and their associated uncertainties with those of M23 for all 43 stars in the sample. As indicated in the legend, the star and circle symbols represent the {\it Kepler} and TESS targets, respectively. The standardised residuals in the left column show the normalised differences between the two determinations of $h_1'$ (top panel) and $h_2'$ (bottom panel) as a function of the impact parameter. The dashed horizontal lines marked at zero correspond to perfect agreement. The right column presents the ratios of our and M23 uncertainties in $h_1'$ (top panel) and $h_2'$ (bottom panel) as a function of the impact parameter. The dashed horizontal lines marked at 1 correspond to the same uncertainties found in both studies.}
    \label{fig:h1ph2p_KV_PM}
\end{figure*}

\begin{figure*}
	\includegraphics[width=\textwidth]{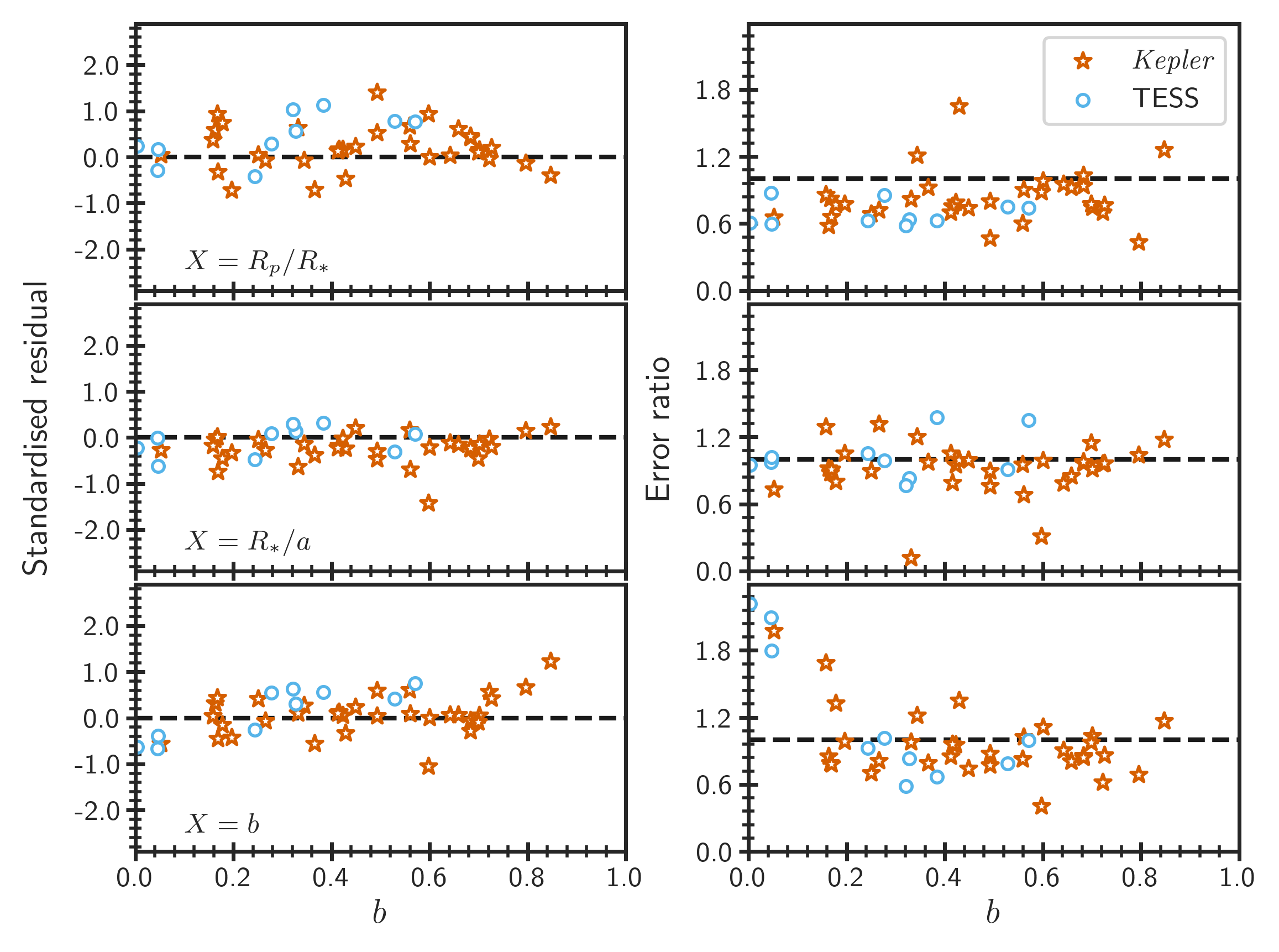}
    \caption{Same as Figure~\ref{fig:h1ph2p_KV_PM} except now we compare our measured $R_p/R_*$, $R_*/a$, and $b$ and their associated uncertainties with those of M23 for all 43 stars in the sample.}
    \label{fig:RpRsb_KV_PM}
\end{figure*}  

\begin{figure*}
	\includegraphics[width=\textwidth]{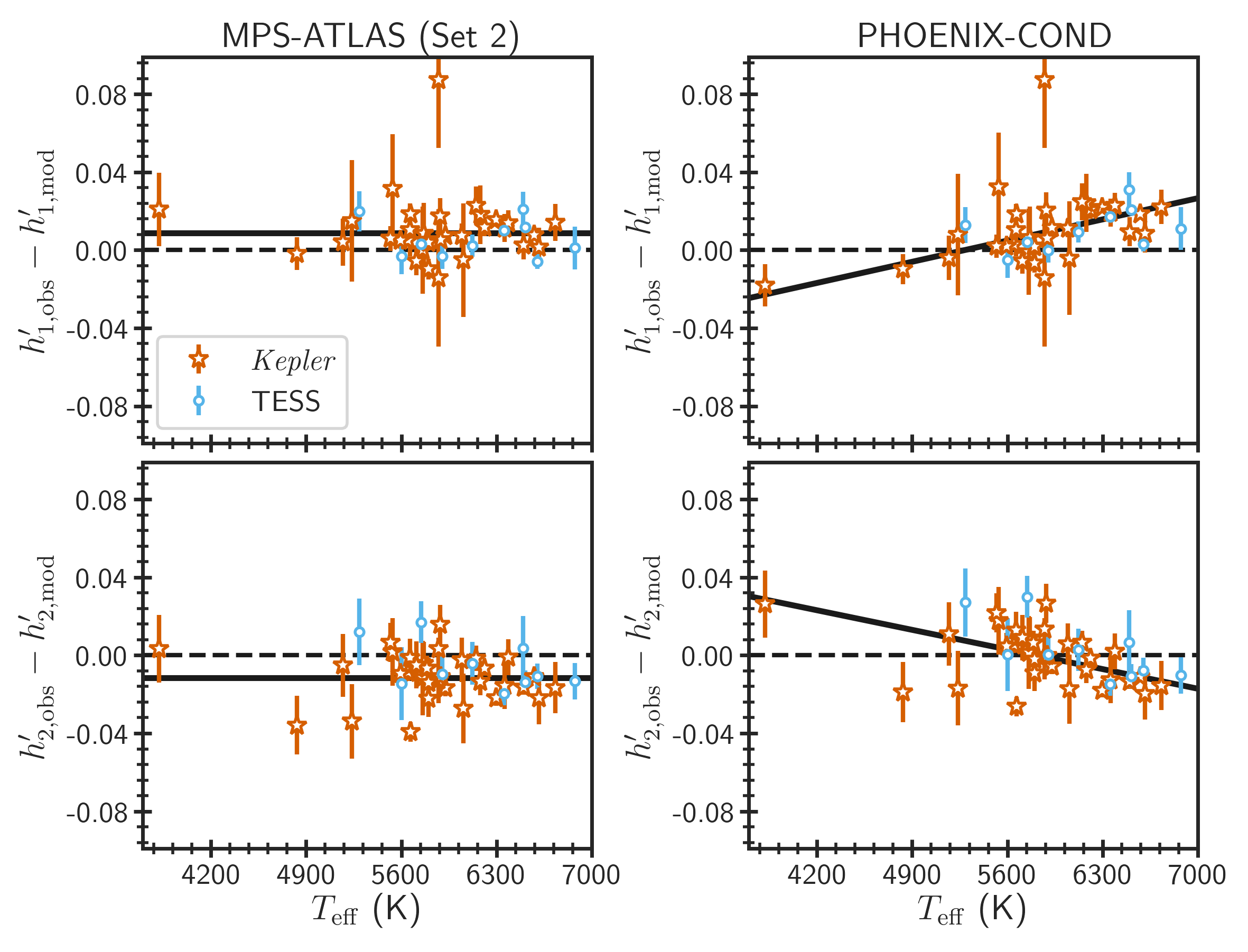}
    \caption{Comparisons of our measured $h_1'$ (top row) and $h_2'$ (bottom row) with the predictions of the MPS-ATLAS (left column) and PHOENIX-COND (right column) models for all the stars in the sample. As indicated in the legend, the star and circle symbols represent the {\it Kepler} and TESS targets, respectively. The dashed horizontal lines marked at zero correspond to the perfect agreement. The solid horizontal lines in the left panels represent the offsets (weighted mean of the residuals), while the solid lines in the right panels show the weighted linear least square fits.}
    \label{fig:h1ph2p_teff_KV_models}
\end{figure*}

\subsection{Analysis of the {\it Kepler} and TESS targets}
\label{subsec:m23}
Having performed various tests to ensure the precision and accuracy of our method, we now proceed to analyse the data of other stars in our sample with $\lambda = 0.2$. Table~\ref{tab:fit_params} lists the resulting parameters along with the reduced chi-square for all 43 stars in our sample. Note that the inferred parameters do not account for the tidal deformation of the planet \citep{burt14,corr14}. To compare our results with M23, we define the standardised residual and error ratio as,   
\begin{equation}
    {\rm Standardised \ residual} = \frac{X_{\rm V} - X_{\rm M}}{\sqrt{\sigma_{\rm V}^2 + \sigma_{\rm M}^2}},
    \label{eq:sresidual}
\end{equation}
and
\begin{equation}
    {\rm Error \ ratio} = \frac{\sigma_{\rm V}}{\sigma_{\rm M}},
    \label{eq:eratio}
\end{equation}
where $X$ and $\sigma$ refer to a specific quantity being compared and its associated uncertainty, respectively. The subscripts V and M indicate determinations from this work and M23, respectively. As we can see in the left panels of Figure~\ref{fig:h1ph2p_KV_PM}, the two inferences of $h_1'$ and $h_2'$ for all the stars in our sample agree well within 2$\sigma$ (for most stars, the agreement is within 1$\sigma$). Interestingly, as can be seen in the top right panel, our estimates of the uncertainty on $h_1'$ for systems with small $b$ are comparable to those of M23, however our $h_1'$ becomes increasingly more precise (by a factor of up to 2) for systems with larger $b$. This behaviour is somewhat expected. Note that the observed transit light curves for the systems with zero or small $b$ carry complete information about the CLV of the intensity. On top of this, if the photometric precision is good enough, the LD model is well constrained by the data themselves. For such targets, fitting with and without regularisation should both provide similar results. On the other hand, if $b$ is large or the photometric precision is poor, regularisation helps significantly in constraining the LD model and leads to better precision on the inferences. From the bottom right panel, we can see that our estimates of $h_2'$ are also more precise than those of M23 (again by a factor of up to 2). 

In Figure~\ref{fig:RpRsb_KV_PM}, we compare our determinations of $R_p/R_*$, $R_*/a$, and $b$ with those of M23. Again, the two inferences agree well as can be seen in the left panels. Clearly, the precision of our $R_p/R_*$ determinations is, on average, better than M23 by a factor of about $1.5$, while the improvements in the precision of $R_*/a$ and $b$ measurements are relatively modest.

\subsection{Comparisons with the model predictions}
\label{subsec:models}
The 1D models of stellar atmospheres can be used to compute the CLV of the specific intensity as a function of the stellar parameters. We shall compare our measured $h_1'$ and $h_2'$ with the LD profiles calculated by \citet{clar18} based on the PHOENIX-COND models \citep{huss13} as well as by \citet{kost22} based on the MPS-ATLAS models \citep{witz21}. \citet{kost22} provides two sets of LD profiles; `Set 1’ uses atmospheric models computed with the \citet{grev98} solar metallicity mixture and a fixed value of the mixing-length parameter, while `Set 2' uses models calculated with the \citet{aspl09} mixture and a variable mixing-length parameter (which depends on the effective temperature). We shall compare our results with their Set 2 model predictions. The observed values of $T_{\rm eff}$, $\log g$ and $[{\rm Fe}/{\rm H}]$ of a star can be used to compute the corresponding predicted values of $h_1'$ and $h_2'$ through interpolation in a grid of model LD profiles. The uncertainties in the predicted $h_1'$ and $h_2'$ due to the observational errors in $T_{\rm eff}$, $\log g$ and $[{\rm Fe}/{\rm H}]$ are calculated using Monte Carlo simulations. We refer the reader to M23 for further details on how the model predictions of $h_1'$ and $h_2'$ for all 43 stars were computed using the MPS-ATLAS and PHOENIX-COND atmospheric models. 

\begin{figure}
	\includegraphics[width=\columnwidth]{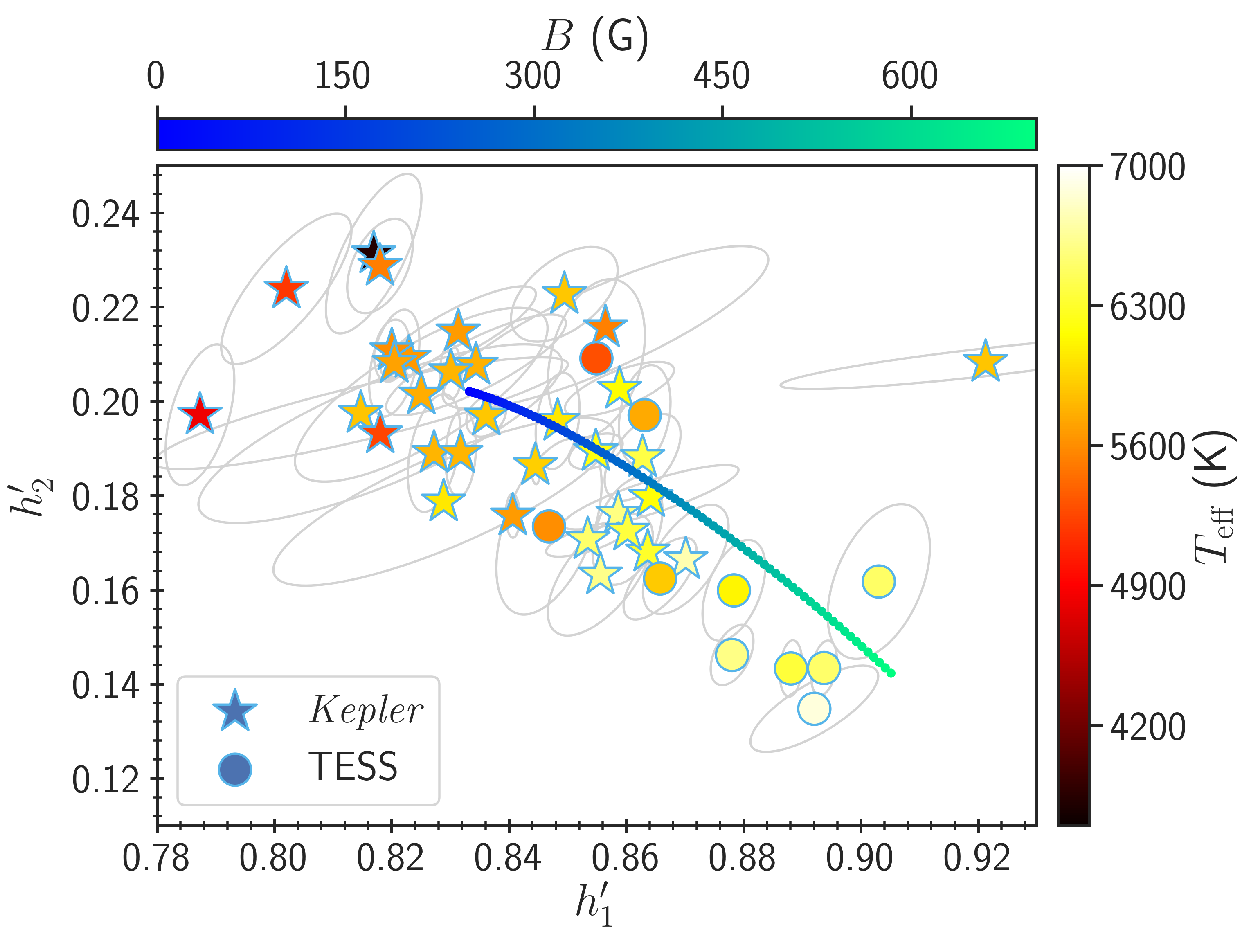}
    \caption{Comparison between the observed limb darkening and the corresponding MHD model predictions in the $h_1' - h_2'$ plane. As indicated in the legend, the star and circle symbols represent the {\it Kepler} and TESS targets, respectively. The ellipses around these symbols highlight the 1$\sigma$ confidence regions. The associated color indicates the effective temperature of the star. The curve is a cubic function of the magnetic field, which was fitted to the results of the MHD models normalised to the solar conditions (see the text). The color of the curve shows the strength of the magnetic field.}
    \label{fig:h1ph2p_Teff_KV}
\end{figure}

Figure~\ref{fig:h1ph2p_teff_KV_models} shows comparisons of the observed $h_1'$ and $h_2'$ with the model predictions. For the PHOENIX-COND models in the right panels, the $h_1'$ and $h_2'$ residuals show linear trends as a function of $T_{\rm eff}$. This was also noted by M23. We fitted straight lines to the residuals and found,
\begin{eqnarray}
    \Delta h_1' &=& (0.016\pm0.002) X + (0.011\pm0.001),\\
    \Delta h_2' &=& (-0.014\pm0.003) X - (0.003\pm0.002),
\end{eqnarray}
where $X = (T_{\rm eff} - 6000) / 1000$. There are no such visual trends in the $h_1'$ and $h_2'$ residuals in the left panels for the MPS-ATLAS models. In this case, an attempt to fit straight lines to the residuals results in slopes that are consistent with zero. However, as can be clearly seen, the models on average underestimate $h_1'$ and overestimate $h_2'$. The offsets (weighted mean of the residuals) for $h_1'$ and $h_2'$ are $0.009 \pm 0.001$ and $- 0.011 \pm 0.001$, respectively. Apart from these discrepancies, if we look at the differences between the observed and model-predicted values of $h_1'$ and $h_2'$ for individual stars, there are several cases of highly significant deviations. Given the systematic uncertainties in $h_1'$ and $h_2'$ are much smaller (see Figure~\ref{fig:simulation}), we can confidently say that both MPS-ATLAS and PHOENIX-COND models have shortcomings. These 1D models neglect the magnetic field, which is known to play an important role and could be one of the reasons behind the above discrepancies.  

Recent 3D MHD simulations of the stellar atmosphere show that the CLVs indeed depend on the surface magnetic field, $B$. Using the CO$^5$BOLD code \citep{frey12}, \citet[][hereafter L23]{ludw23} constructed a set of nine MHD models assuming solar atmospheric conditions with $B$ varying from 0 to 2400 G. The chemical composition and surface gravity of all the simulations were kept fixed at the corresponding solar values; however, the resulting $T_{\rm eff}$ does not necessarily correspond to the solar value and, in fact, depends on $B$. To isolate the effect of the magnetic field, L23 developed a normalisation procedure that modifies the values of $h_1'$ and $h_2'$ for a star with non-solar atmospheric parameters such that they correspond to the solar conditions. For each simulation, they obtained three different sets of LD parameters: one from the light-curve fitting, the second resulting from the normalisation of the first set (corrects for differing $T_{\rm eff}$ only since all other parameters already correspond to solar conditions), and the last from direct calculations based on the underlying Claret 4-parameter law. For further details on the MHD models, various methods to calculate the LD parameters, and the normalisation procedure, we refer the reader to L23.   

In Figure~\ref{fig:h1ph2p_Teff_KV}, we show our measurements in the $h_1' - h_2'$ plane. There is a clear trend in the diagram; $h_2'$ decreases as a function of $h_1'$. Furthermore, we can see the dependence of the LD parameters on $T_{\rm eff}$. For comparison, we also plot the L23 predictions in the form of a curve that was fitted to the results of the MHD simulations normalised to the solar conditions. Note the apparent near-degeneracy between $T_{\rm eff}$ and $B$ (both increase as we go from the upper left to the lower right in the figure). This is clearly an unfortunate situation for the inference of stellar magnetic field from LD observations because this requires an accurate measurement of $T_{\rm eff}$. The simulations reproduce the observational trend reasonably well; however, it is interesting to note that a larger fraction of the data points lie below the model curve. 

\begin{figure*}
	\includegraphics[width=\textwidth]{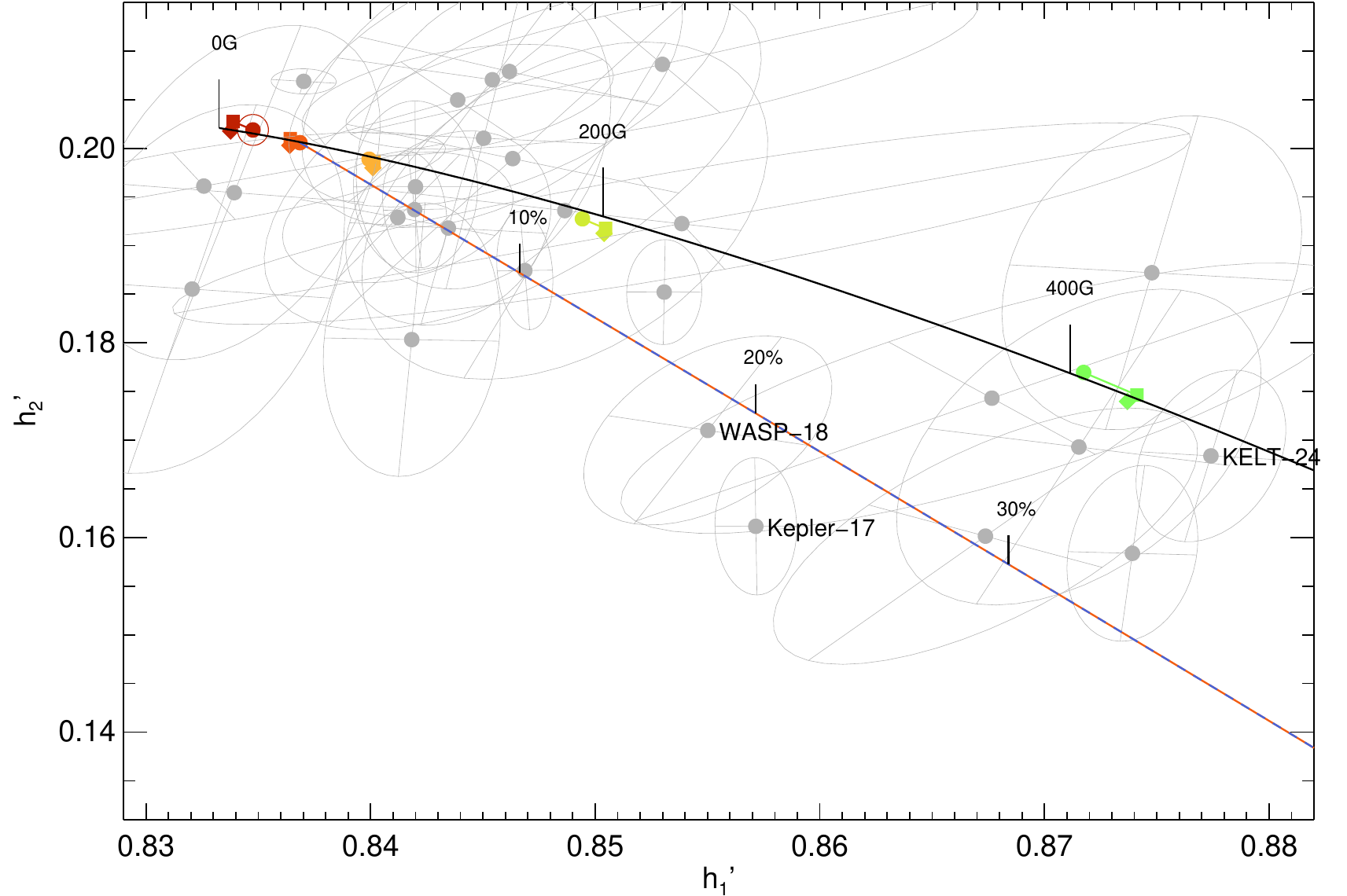}
    \caption{Comparison between the normalised observed data and corresponding MHD model predictions in the $h_1' - h_2'$ plane. The grey filled circles represent stars, and the ellipses around them highlight the 1$\sigma$ confidence region. The colourful square, circle, and diamond symbols depict the results of the MHD simulations as obtained from the light-curve fitting, normalisation procedure, and direct calculations based on the underlying Claret 4-parameter law, respectively. The model points before and after normalisation are connected by a line. The solid black curve is a cubic function of the magnetic field, which was fitted to the results of the MHD models normalised to the solar conditions. The colorful dashed curve illustrates the outcome of an additive mixing of models with magnetic fields of 50 and 1600 G. The labeled percentages give the area fraction contributed by the simulation with a 1600 G magnetic field. The best-guess solar position as obtained from the simulation with 0 G magnetic field after normalisation is marked by the solar symbol.}
    \label{fig:h1ph2p_HL}
\end{figure*}

To study the effect of magnetic field on the LD and also perform a cleaner comparison between the observation and simulations, we eliminate the dependence of the measured $h_1'$ and $h_2'$ on atmospheric parameters by normalising them to the solar conditions following the procedure of L23. The normalised observed data and simulation curves are shown in Figure~\ref{fig:h1ph2p_HL}. Note that a fraction of stars are absent in the diagram as a result of their large error regions. In this diagram, if the stars had zero magnetic field and the measurements had no errors, then all the observed data points would fall to the solar position. Clearly, We see a trend that is qualitatively similar to what is predicted by the MHD simulations. However, as noted earlier, a larger fraction of stars falls below the model curve. This is likely a result of our choice of homogeneous magnetic field in the simulations, which is unrealistic. We know that the morphology of the stellar surface magnetic field is more complex, with much higher field strengths in the active regions (like stellar spots) than in the quiescent regions. As we can see in the figure, the curve corresponding to an additive mixing of two simulations with magnetic fields of 50 and 1600 G fits the data trend better. 

From the MHD simulations, we expect that stars with increasingly large homogeneous magnetic fields follow a smooth trend extending from the upper left to the lower right in Figure~\ref{fig:h1ph2p_HL}. The results of mixing two simulations suggest that the localised high $B$ regions, such as star spots on top of the homogeneous background, lead to scatters around the smooth trend. This is indeed what we see in the observed data. Among the stars on the lower right, we observe additional scatter in the light curve of Kepler-17 during the transit due to random occultation of active regions by the planet (see also, M23), showing its elevated magnetic activity. However, for the rest, this signature of magnetic activity in the light curve is not observed. This does not necessarily mean that these stars have low magnetic fields. Note that if a star is near the minimum of its stellar cycle or the transit occurs at high latitudes (where we do not expect to have as many stellar spots even during the maximum), the planet may not occult any spots of significant sizes. The case of WASP-18 is particularly interesting in this regard. It is expected to have a relatively high magnetic field for two reasons: (1) its position in Figure~\ref{fig:h1ph2p_HL}; and (2) its fast rotation and young age \citep{hell09}. However, it neither shows additional scatter in the light curve nor has been detected in X-ray observations \citep{pill14}, implying a very low level of magnetic activity. A possible explanation could be that this star is going through a grand minimum of activity like the ones observed for the Sun \citep[see e.g.][]{bisw23}. Recently, \citet{lanz24} proposed an alternative explanation in which a massive close-by planet (such as the one WASP-18 has) can tidally induce turbulence which can inhibit the emergence of magnetic flux tubes responsible for the formation of the photospheric star spots. \citet{pill23} observed KELT-24 and, as we expect from Figure~\ref{fig:h1ph2p_HL}, they found it to be an active star showing X-ray emission.

\section{Conclusions}
\label{sec:conc}
We developed a novel Python-based Bayesian Exoplanet Transit Modelling tool, BETMpy\footnote{https://github.com/kuldeepv89/BETMpy}, to fit the model transit light curves to the observed data and accurately infer planet and orbital properties and characterise stellar limb darkening. Although the method used \textsc{BATMAN} transit models with the complex nonlinear LD law, its complexity was appropriately tuned during the fitting process through a second derivative regularisation. To estimate the regularisation parameter, we analysed the observed transit light curve of Kepler-5 with different values of this parameter. We found that $\lambda = 0.2$ provides a good fit to the data and also leads to a tightly constrained brightness distribution of the projected stellar disk and precise inferences of the LD parameters $h_1'$ and $h_2'$. 

To test the accuracy of the inferred parameters, we applied our technique to a set of 900 simulated transit light curves with a wide range of impact parameters computed using the \textsc{BATMAN} package. We found that our technique recovers the parameters reasonably well, with the differences between the inferred properties and the corresponding true values generally lying within the 1$\sigma$ statistical uncertainties. This exercise reassures the accuracy of our inferences. It was also shown that our uncertainty estimates from the posterior probability distribution are reliable and consistent with Monte Carlo uncertainties.

We used our validated method to analyse the transit light curves of a sample of 43 systems observed by the NASA {\it Kepler} and TESS missions. We found that our measurements are consistent with the latest inferences of M23 while being significantly more precise. In particular, the errorbars on the LD parameters $h_1'$ and $h_2'$ are smaller than those obtained in M23 by a factor of up to 2. The precision of our inferred planet-to-star radius ratio is, on average, better than that of M23 by a factor of 1.5. Since M23, with poorer measurement precision, demonstrated that the uncertainties in the LD parameters are not the limiting factor for the precision of planet radii, our more precise LD measurements can be confidently used to accurately determine them during the PLATO mission.

We compared our LD measurements with the predictions of the MPS-ATLAS and PHOENIX-COND models and found that these 1D non-magnetic models fail to reproduce the observations. As seen in the $h_1' - h_2'$ diagram, the predictions of the 3D MHD simulations follow the same trend as the observations; however, there is still a noticeable discrepancy, as relatively more data points lie below the model curve. To remove the impact of the atmospheric parameters from the limb darkening and perform a more consistent comparison between the model and observation, we normalised the observed data to solar conditions following the procedure of L23. We found that the above discrepancy persists. We attributed the origin of this problem to our unrealistic choice of the homogeneous magnetic field in the simulations. We showed that a physically more realistic model corresponding to an additive mixing of two simulations with magnetic fields of 50 and 1600 G fits the data better. Our precise measurements of the LD, together with MHD simulations, confirm that Kepler-17, WASP-18, and KELT-24 indeed have relatively high magnetic fields ($> 200$ G). This study demonstrates that we can potentially estimate the stellar surface magnetic field by calibrating the observed LD against the predictions of realistic MHD simulations of varying magnetic fields.

\section*{Acknowledgements}
This research was supported by the Munich Institute for Astro-, Particle and BioPhysics (MIAPbP) which is funded by the Deutsche Forschungsgemeinschaft (DFG, German Research Foundation) under Germany´s Excellence Strategy – EXC-2094 – 390783311. The support and the resources provided by PARAM Shivay Facility under the National Supercomputing Mission, Government of India at the Indian Institute of Technology, Varanasi are gratefully acknowledged. We thank the anonymous referee for helpful comments.

\section*{Data Availability}
The data underlying this article will be shared on reasonable request to the corresponding author. The code for the Bayesian exoplanet transit modelling with an example case (Kepler-5) is available on https://github.com/kuldeepv89/BETMpy.


\appendix

\section{Sensitivity to regularisation parameter}
\label{app:lambda}

\begin{figure*}
	\includegraphics[width=0.80\textwidth]{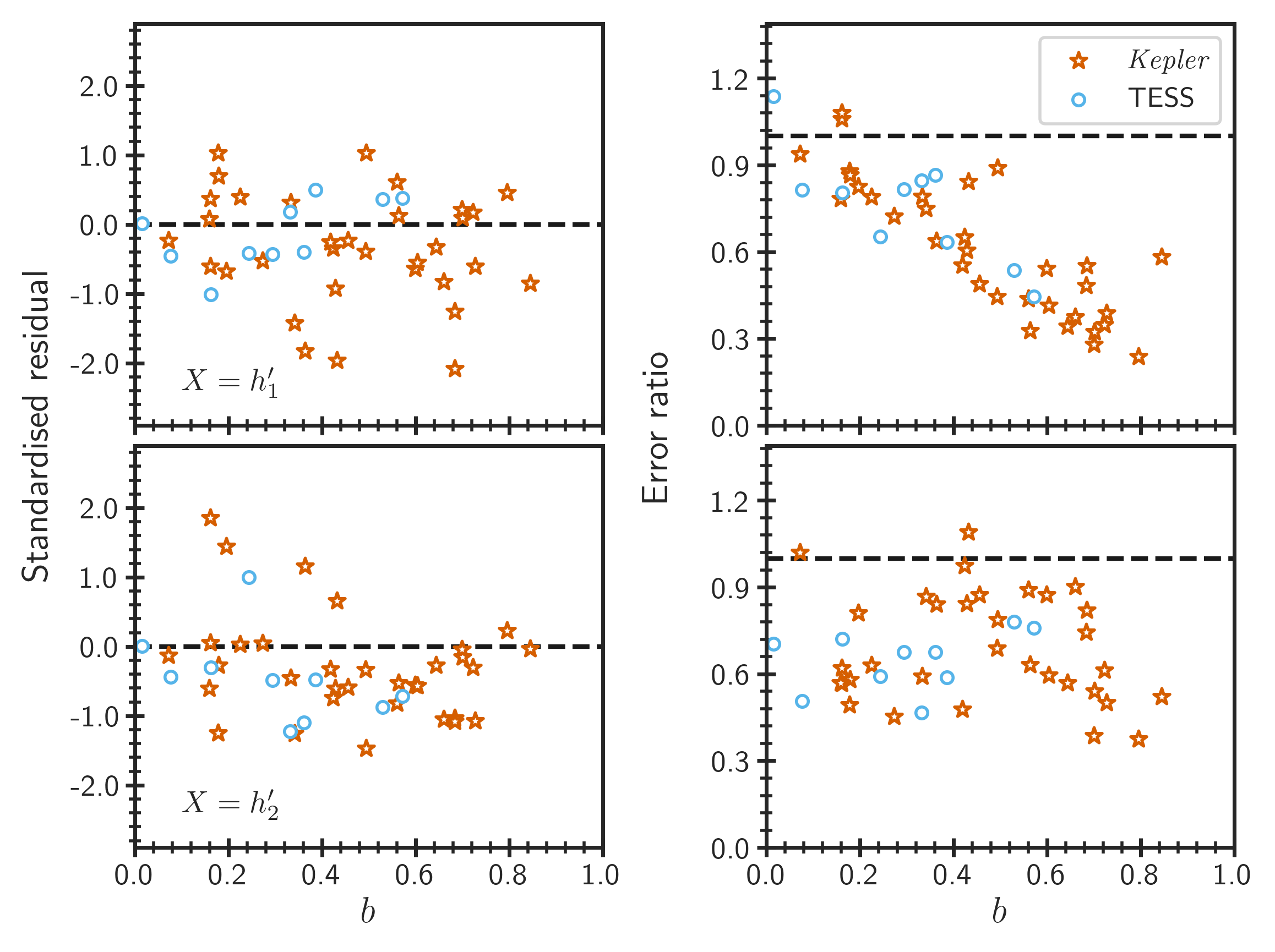}
    \caption{Same as Figure~\ref{fig:h1ph2p_KV_PM} except the analysis was carried out with $\lambda=0.5$ (instead of 0.2).}
    \label{afig:h1ph2p_KV_PM_5}
\end{figure*}

\begin{figure*}
	\includegraphics[width=0.80\textwidth]{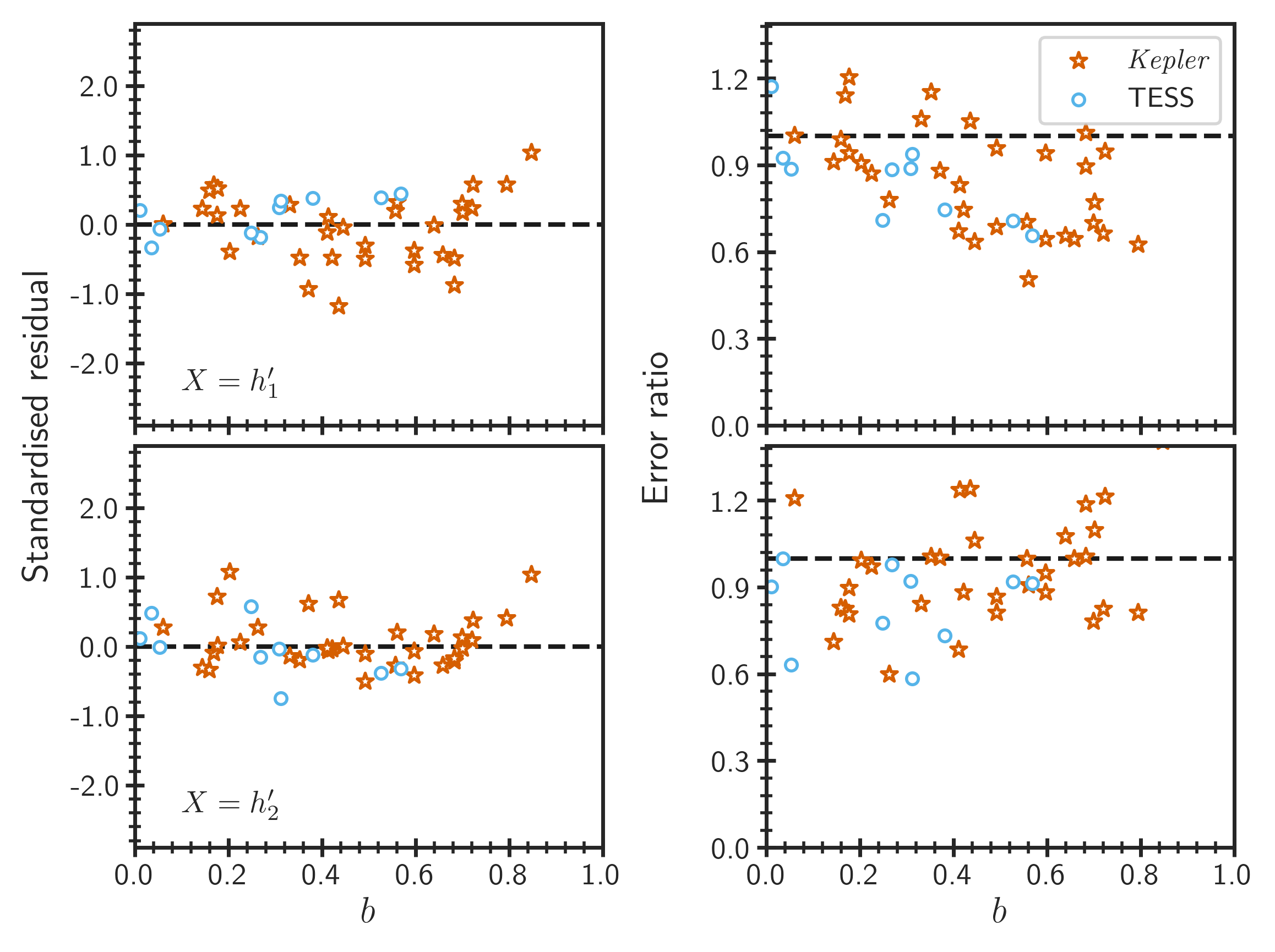}
    \caption{Same as Figure~\ref{fig:h1ph2p_KV_PM} except the analysis was carried out with $\lambda=0.1$ (instead of 0.2).}
    \label{afig:h1ph2p_KV_PM_1}
\end{figure*}

In this section, we show the results obtained with the values of the regularisation parameter in the neighbourhood of the reference value (0.2). In Figure~\ref{afig:h1ph2p_KV_PM_5}, we compare our estimates of $h_1'$, $h_2'$ and their associated uncertainties found with $\lambda = 0.5$ with the corresponding determinations of M23. In the right panels, it is clear that the measurement precision has improved; however, we notice a small bias in the left panels. In Figure~\ref{afig:h1ph2p_KV_PM_1}, we show the same, but now the analysis was performed with $\lambda = 0.1$. With this choice of $\lambda$, we almost completely remove the bias, but significantly lose precision. This behaviour is clearly expected from the discussion in the body text. 

\bsp
\label{lastpage}
\end{document}